\documentclass[prd,nofootinbib,preprint,superscriptaddress]{revtex4}

\usepackage{amsmath, amssymb, amsthm, graphicx, epsfig, fancyhdr,epsfig, slashed}

\usepackage{tikzsymbols}
\usepackage{natbib}
\usepackage{float}

\usepackage{tikz,xcolor,hyperref}

\usepackage{mathtools}

\definecolor{lime}{HTML}{A6CE39}
\DeclareRobustCommand{\orcidicon}{
	\begin{tikzpicture}
	\draw[lime, fill=lime] (0,0) 
	circle [radius=0.2] 
	node[white] {{\fontfamily{qag}\selectfont \tiny ID}};
	\draw[white, fill=white] (-0.0625,0.095) 
	circle [radius=0.007];
	\end{tikzpicture}
	\hspace{-2mm}
}

\foreach \x in {A, ..., Z}{\expandafter\xdef\csname orcid\x\endcsname{\noexpand\href{https://orcid.org/\csname orcidauthor\x\endcsname}
			{\noexpand\orcidicon}}
}

\newcommand{\be}{\begin{equation}}
\newcommand{\ee}{\end{equation}}
\newcommand{\bea}{\begin{eqnarray}}
\newcommand{\eea}{\end{eqnarray}}

\newcommand{\eps}{\varepsilon}
\newcommand{\pfrac}[2]{\left(\frac{#1}{#2}\right)}
\newcommand{\GeV}{{\rm\,GeV}}
\newcommand{\sn}{\mathop{\rm sn}\nolimits}

\newcommand{\bbbone}{\hbox{\rm 1\kern-3pt l}}
\newcommand{\slp}{p\kern-5pt/}

\newcommand{\tr}{\mathop{\rm tr}\nolimits}

\begin{document}

\title{On a novel evalutation of the hadronic contribution \\
  to the muon's $g-2$ from QCD}

\author{Marco Frasca\orcidA{}}
\email{marcofrasca@mclink.it}
\affiliation{Rome, Italy}

\author{Anish Ghoshal\orcidB{}}
\email{anish.ghoshal@roma2.infn.it}
\affiliation{Institute of Theoretical Physics, Faculty of Physics, University of Warsaw,ul.  Pasteura 5, 02-093 Warsaw, Poland}
\affiliation{INFN - Sezione Roma ``Tor Vergata'',
  Via della Ricerca Scientifica 1, 00133, Roma, Italy}

\author{Stefan Groote\orcidC{}}
\email{stefan.groote@ut.ee}
\affiliation{F\"u\"usika Instituut, Tartu Ulikool,
  W.~Ostwaldi 1, EE-50411 Tartu, Estonia}

\begin{abstract}
\textit{We evaluate the hadronic contribution to the $g-2$ of the
muon by deriving the low-energy limit of quantum chromodynamics (QCD) and
computing in this way the hadronic vacuum polarization. The low-energy limit is a
non-local Nambu--Jona-Lasinio (NJL) model that has all the parameters fixed from QCD, and the only experimental input used is the
confinement scale that is known from measurements of hadronic physics. Our estimations provide a novel analytical alternative to the current lattice computations and we find that our result is close to the similar computation performed from experimental data. 
We also comment on how this analytical approach technique, in general, may provide prospective estimates for hadronic computations from dark sectors and its implication in BSM model-building in future.}
\end{abstract}

\maketitle

\section{Introduction}

Since the original computation for the electron from first principles~\cite{Commins:2012nc}, originating from Dirac equation, 
the lepton anomalous magnetic moments continue to be very important observables for
precision tests of the Standard Model (SM)~\cite{Kinoshita:1990nb}. 
Recent data seems to indicate a tension with the theoretical prediction for the 
anomalous magnetic moment of the muon with the recent experimental value for the
anomalous magnetic moment of the muon is~\cite{Bennett:2006fi,Muong-2:2021ojo}
\begin{equation}
g_\mu/2=1+a_\mu=1.001\,165\,920\,8(6).
\end{equation}
The Particle Data Group (PDG) gives an updated value for the muon anomaly
in the form~\cite{PDG}
\begin{equation}
a_\mu^{\rm exp} = 116\,592\,091(54)(33)\times 10^{-11}.
\end{equation}
%
This precision clearly is a challenge for the 
theoretical side to increase the precision of the prediction~\cite{Muong-2:2021ojo}

The theoretical results for the muon anomalous magnetic moment in the SM are 
traditionally represented as a sum of three parts,
\begin{equation}
\label{eq:a-SM}
  a_\mu^{\rm SM}=a_\mu^{\rm QED}+a_\mu^{\rm EW}+a_\mu^{\rm had}
\end{equation}
with $a_\mu^{\rm QED}$, $a_\mu^{\rm EW}$ being the leptonic and electroweak parts,
respectively, and $a_\mu^{\rm had}$ is the contribution involving the electromagnetic 
currents of quarks. 

The leptonic part is computed in perturbation theory and reads~\cite{PDG}
\begin{equation}
a_\mu^{\rm QED}=116\,584\,718.95(0.08)\times 10^{-11}.
\end{equation}
%

The electroweak part is known to two loops and reads~\cite{PDG} 
\begin{equation}
a_\mu^{\rm EW} = 153.6(1.0)\times 10^{-11}\, .
\end{equation}
%

The hadronic part $a_\mu^{\rm had}$ in the SM is related to quark contributions
to the electromagnetic currents. 

The current
total SM prediction reads~\cite{PDG}
\begin{equation}
a_\mu^{\rm SM}=116\,591\,823(1)(34)(26)\times 10^{-11}.
\end{equation}
The difference
\begin{equation}
\Delta a_\mu = a_\mu^{\rm exp}-a_\mu^{\rm SM}=268(63)(43)\times 10^{-11}
\end{equation}
could be due to new unknown physics beyond the SM, but it is not statistically
significant off yet~\cite{Aoyama:2020ynm}, the
main idea behind this being that contributions from unknown virtual
particles not part of the SM might enter the calculations.

In general, theoretical estimates are very precise for what one should expect
from quantum electrodynamics (QED), but fall short in the case of the hadronic
contributions, due to the known difficulties to treat quantum chromodynamics
(QCD) at low energies. The general accepted technique is to use experimental
results from e$^+$e$^-$ scattering into hadrons measured in collider
experiments~\cite{Aoyama:2020ynm}. Two key ingredients of this contribution
are the hadronic vacuum polarization (HVP) and the high-order hadronic
light-by-light scattering (HLbL). Of these two contributions, the former is
the most critical one, due to the current inability to compute this
contribution starting right from the Lagrangian of QCD.
Some
recent evaluation of
the HVP from experimental data is given in Ref.~\cite{Keshavarzi:2018mgv,Colangelo:2020lcg,Davier:2019can} for
the $\pi\pi$ part which is the most relevant contribution. The value of the HVP
contribution determines in a critical way whether there is room for Beyond-Standard Model (BSM)
physics or not in the context of observed values for muon $g-2$.

The only independent technique for calculations in QCD is by using large
computer facilities to solve the equations, a technique known as lattice QCD.
Still, the Muon $g-2$ Theory Initiative~\cite{Aoyama:2020ynm} decided to not
use this technique, as there are large differences between the results of
different collaborations, disclosing the technique not to be yet trustworthy.
For instance, the Budapest--Marseille--Wuppertal (BMW) Collaboration has put
forward their latest results~\cite{Borsanyi:2020mff}, showing that the HVP
correction they obtain moves the ballpark of the muon $g-2$ value back into
the SM field. In turn, this would imply that the technique using
experimental values from the colliders probably underestimates this
contribution.

Working on QCD, one generally makes the use of effective models. However, it is
often unknown if such models could be straightforwardly obtained from the
original Lagrangian. Still, successful results have been obtained from some of
these models. In the very early days of the study of the muon $g-2$ problem,
attempts were made to derive the HVP contribution from such effective models
like for instance the Nambu--Jona-Lasinio (NJL) model~\cite{Klevansky:1992qe},
detailed by \textit{de Rafael} in Ref.~\cite{deRafael:1993za} and more recently \cite{Dorokhov:2016mxa}. However, due to the
large set of undetermined parameters entering in such effective theories, this
kind of approach in this early, primitive stage was abandoned in favor of the
use of experimental data and lattice QCD calculations. 

Inspired by such an approach, in this article we will show how an effective field theory can be derived from QCD, starting directly from the Lagrangian level. The model is a non-local
Nambu--Jona-Lasinio model, having all the parameters properly fixed. A first attempt in this direction was given in Ref.~\cite{Frasca:2019ysi} in order to determine 
the proper low-energy limit of the theory\footnote{Unfortunately, this publication contained a mistake that made the conclusions
unreliable.}. In this work, we fix an error in this publication and show how the effective NJL
model comes out naturally from QCD. Based on these first
principles, we will evaluate the HVP contribution 
to the muon $(g-2)$.


\medskip

\section{Basic equations for NJL Model\label{sec2}}

Our starting point is the well-known QCD Lagrangian
\begin{equation}
{\cal L}_{QCD}=\sum_i\bar q_i(i\gamma^\mu D_\mu+m)q_i
  -\frac14F^{\mu\nu}_aF_{\mu\nu}^a
  -\frac1{2\xi}(\partial_\mu A^\mu_a)(\partial_\nu A^\nu_a)
\end{equation}
with covariant derivative $D_\mu=\partial_\mu+igT_aA_\mu^a$ and the field
strength tensor components $F_{\mu\nu}^a$ defined by
$igT_aF_{\mu\nu}^a=[D_\mu,D_\nu]$. The sum over $i$ is understood to run over
the quark flavors and colors. Throughout this paper we work with the
Minkowskian metric $g^{\mu\nu}=\mbox{\rm diag}(1,-1,-1,-1)$. Calculating the
Euler--Lagrange equations, one obtains
\begin{eqnarray}
0&=&\frac{\partial{\cal L}_{QCD}}{\partial A_\nu^a}
  -\partial_\mu\frac{\partial{\cal L}_{QCD}}{\partial(\partial_\mu A_\nu^a)}
  \ =\nonumber\\
  &=&\partial_\mu(\partial^\mu A^\nu_a-\partial^\nu A^\mu_a)
  +\frac1\xi\partial^\nu(\partial_\mu A^\mu_a)
  +gf_{abc}\partial_\mu(A^\mu_bA^\nu_c)\strut\nonumber\\&&\strut
  +gf_{abc}(\partial^\mu A^\nu_b-\partial^\nu A^\mu_b)A_\mu^c
  +g^2f_{abc}f_{cde}A^\mu_bA^\nu_dA_\mu^e-g\sum_i\bar q_i\gamma^\nu T_aq_i,
  \nonumber\\
0&=&\frac{\partial{\cal L}_{QCD}}{\partial\bar q_i}
  -\partial_\mu\frac{{\cal L}_{QCD}}{\partial(\partial\bar q_i)}
  \ =\ (i\gamma^\mu D_\mu+m)q_i.
\end{eqnarray}
These classical equations of motion are the starting point for a tower of
Dyson--Schwinger equations. In order to study these equations, we use the
method proposed by Bender, Milton and Savage~\cite{Bender:1999ek}, details of
which can found in Refs.~\cite{Frasca:2013kka,Frasca:2013tma,%
Frasca:2015yva,Frasca:2017slg,Frasca:2019ysi}. For the purpose of this
publication we sketch the main steps here,  skipping contributions
from BRST ghosts for simplicity.

Enlarging the Lagrangian of the classical action by adding corresponding
source terms $A_\mu^aJ^\mu_a$, $\bar q_i\eta_i$ and $\bar\eta_iq_i$, one
obtains the exponential of the generating functional. Functional derivatives
of this generating functional lead to the Dyson--Schwinger analogue of the
Euler--Lagrange equations, expressed in terms of Green functions for the
fields. The set of equations in Landau gauge $\xi=0$ we start with is given by 
\begin{eqnarray}
\lefteqn{\partial^2A_\nu^{1a}(x)+gf_{abc}
  \left(\partial^\mu A_{\mu\nu}^{2bc}(x,x)+\partial^\mu A_\mu^{1b}(x)
  A_\nu^{1c}(x)-\partial^\nu A_{\mu\nu}^{2bc}(x,x)
  -\partial_\nu A_\mu^{1b}(x)A_c^{1\mu}(x)\right)\strut}\nonumber\\&&\strut
  +gf_{abc}\partial^\mu A_{\mu\nu}^{2bc}(x,x)
  +gf_{abc}\partial^\mu(A_\mu^{1b}(x)A_\nu^{1c}(x))	
  +g^2f_{abc}f_{cde}\Big(g^{\mu\rho}A_{\mu\nu\rho}^{3bde}(x,x,x)
  \strut\nonumber\\&&\strut
  +A_{\mu\nu}^{2bd}(x,x)A_e^{1\mu}(x)+A_{\nu\rho}^{2eb}(x,x)A_d^{1\rho}(x)
  +A_{\mu\nu}^{2de}(x,x)A_b^{1\mu}(x)+A_b^{1\mu}(x)A_\mu^{1d}(x)A_\nu^{1e}(x)
  \Big)\ =\nonumber\\
  &=&g\sum_i\gamma_\nu T^aq_{ii}^2(x,x)+g\sum_i\bar q_i^1(x)\gamma_\nu
  T^a q_i^1(x),\nonumber\\
\lefteqn{(i\slashed\partial-m_q)q_i^1(x)
  +g\gamma^\mu A_\mu^{1a}(x)T_aq_i^1(x)\ =\ 0,}
\end{eqnarray}
where the one-, two- and three-point Green functions are given by
$A_\mu^{1a}(x)=\langle A_\mu^a(x)\rangle$,
$A_{\mu\nu}^{2ab}(x,y)=\langle A_\mu^a(x)A_\nu^b(y)\rangle$,
$A_{\mu\nu\rho}^{3abc}(x,y,z)=\langle A_\mu^a(x)A_\nu^b(x)A_\rho^c(x)\rangle$,
$q_i^1(x)=\langle q_i(x)\rangle$ and
$q_{ij}^2(x,y)=\langle q_i(x)q_j(y)\rangle$.
The expected solutions can be written in the form
\begin{equation}
A_\nu^{1a}(x)=\eta_\nu^a\phi(x),\qquad
A_{\mu\nu}^{2ab}(x,y)=\left(g_{\mu\nu}
  -\frac{\partial_\mu\partial_\nu}{\partial^2}\right)\delta^{ab}\Delta(x-y),
\end{equation}
where $\eta_\mu^a$ are the coefficients of the polarization vector with
$\eta_\mu^a\eta^\mu_b=\delta_{ab}$, $\phi(x)$ is a scalar field and
$\Delta(x-y)$ is the propagator of the scalar field. The three-point function
can be set to zero. For the one-point functions we obtain
\begin{eqnarray}\label{Dirac}
\eta_\nu^a\partial^2\phi(x)+2N_cg^2\Delta(0)\eta_\nu^a\phi(x)
  +N_cg^2\eta_\nu^a\phi^3(x)&=&g\sum_i\gamma_\nu T_aq_{ii}^2(x,x)
  +g\sum_i\bar q_i^1(x)\gamma_\nu T^aq_i^1(x),\nonumber\\
(i\gamma^\mu\partial_\mu-m_q)q_i^1(x)+g\gamma^\mu\eta_\mu^aT_a\phi(x)q_i^1(x)
  &=&0.
\end{eqnarray}
Using $\eta_\mu^a\eta^\mu_a=N_c^2-1$ and $\sum_iq_{ii}^2(x,x)=N_cN_fS(0)$, the
first differential equation~(\ref{Dirac}) takes the form
\begin{equation}
\partial^2\phi(x)+2N_cg^2\Delta(0)\phi(x)+N_cg^2\phi^3(x)
  =\frac{g}{N_c^2-1}\left[N_cN_f\gamma^\nu\eta_\nu^aT_aS(0)
  +\sum_i\bar q_i^1(x)\gamma^\nu\eta_\nu^aT_aq_i^1(x)\right].
\end{equation}
In the 't Hooft limit $N_c\to\infty$, $\lambda:=N_cg^2\gg 1$ finite but large,
this set of equations yields a Nambu--Jona-Lasinio model in a straightforward
way. Indeed, for this case, we can perform a perturbation series expansion
$\phi(x)=\phi_0(x)+\phi_1(x)+O(g^2)$ in $g$, obtaining at leading order
\begin{equation}
\partial^2\phi_0(x)+2\lambda\Delta(0)\phi_0(x)+\lambda\phi_0^3(x)=0,
\end{equation}
while the next-to-leading order yields
\begin{eqnarray}\label{NLO}
\lefteqn{\partial^2\phi_1(x)+2\lambda\Delta(0)\phi_1(x)
  +3\lambda\phi_0^2(x)\phi_1(x)\ =}\nonumber\\
  &=&\frac{g}{N_c^2-1}\left[N_cN_f\gamma^\nu\eta_\nu^aT_aS(0)
  +\sum_i\bar q_i^1(x)\gamma^\nu\eta_\nu^aT_aq_i^1(x)\right].
\end{eqnarray}

\subsection{Zeroth order solution and Green function}
Note that $\Delta(0)$ is a constant. Therefore, $m^2=2\lambda\Delta(0)$ can be
considered as the mass square of the scalar field. The leading order
differential equation $\partial^2\phi_0(x)+m^2\phi_0(x)+\lambda\phi_0^3(x)=0$
is nonlinear, but a solution in terms of Jacobi's elliptic
functions exists,
\begin{equation}
\phi_0(x)=\sqrt{\frac{2(p^2-m^2)}\lambda}
  \sn\left(p\cdot x+\theta\,|\,\kappa\right)
\end{equation}
with
\begin{equation}
p^2=\frac12\left(\sqrt{m^4+2\lambda\mu^4}+m^2\right)\quad\mbox{and}\quad
\kappa=\frac{m^2-p^2}{p^2},
\end{equation}
where $\mu$ and $\theta$ are integration constants. $\sn(z|\kappa)$ is Jacobi's
elliptic function of the first kind. 
Given this solution,
the second differential equation can be solved by noting that
\begin{equation}
\left(\partial^2+m^2+3\lambda\phi_0^2(x)\right)\Delta(x-y)=i\delta^4(x-y)
\end{equation}
is solved by a Green function written in momentum space as
\begin{equation}\label{Delta0mom}
\tilde\Delta(p)=\hat Z(p^2,m^2)\frac{2\pi^3}{K^3(\kappa)}
  \sum_{n=0}^\infty(-1)^n\frac{e^{-(n+1/2)\varphi(\kappa)}}
  {1-e^{-(2n+1)\varphi(\kappa)}}\frac{(2n+1)^2}{p^2-m_n^2+i\epsilon}
\end{equation}
with
\begin{equation}
\varphi(\kappa)=\frac{K^*(\kappa)}{K(\kappa)}\pi,\qquad K^*(z)=K(1-z)
\end{equation}
and
\begin{equation}
\hat Z(p^2,m^2)=\frac{2p^8\sqrt{p^2}(p^6+2p^4m^2-3p^2m^4+m^6)}{\sqrt{2p^2-m^2}
  (2p^{12}(2p^2-m^2)-5p^2m^8(2p^2-m^2)(p^2-m^2)-m^{14})}.
\end{equation}
The mass spectrum is given by
\begin{equation}
\label{eq:spec}
m_n=\frac{(2n+1)\pi}{2K(\kappa)}\sqrt{2p^2}=:(2n+1)m_G(\kappa).
\end{equation}
At this point the circle for the mass of the scalar field is closed. Inserting
back the Fourier transform of the propagator~(\ref{Delta0mom}) into
$m^2=2\lambda\Delta(0)$ results in
\begin{equation}\label{m2}
m^2=2\lambda\int\frac{d^4p}{(2\pi)^4}\hat Z(p^2,m^2)\frac{2\pi^3}{K^3(\kappa)}
  \sum_{n=0}^\infty(-1)^n\frac{e^{-(n+1/2)\varphi(\kappa)}}{1-e^{-(2n+1)
  \varphi(\kappa)}}\frac{(2n+1)^2}{p^2-(2n+1)^2m_G^2(\kappa)+i\epsilon}.
\end{equation}
This self-consistency equation provides the proper spectrum of a Yang-Mills
theory with no fermions~\cite{Frasca:2017slg}, in very close agreement with
lattice data.

\subsection{First order solution}
The convolution of the propagator $\Delta$ with the right hand side of
Eq.~(\ref{NLO}) leads to
\begin{eqnarray}\label{eq:phi_1}
  \phi_1(x)=\frac{g}{N_c^2-1}\int d^4y\Delta(x-y)
  \left[N_cN_f\gamma^\nu\eta_\nu^aT_aS(0)
  +\sum_i{\bar q}_i^1(y)\gamma^\nu\eta_\nu^aT_aq_i^1(y)\right].
\end{eqnarray}
The first term renormalizes the fermion mass and can taken to be zero by
choosing the renormalization condition $S(0)=0$. The second term yields a
Nambu--Jona-Lasinio (NJL) interaction in the equation of motion of the quark. 

Inserting $\phi(x)$ into the Dirac equation~(\ref{Dirac}), in the 't Hooft
limit the term $\phi_0$ is negligible small compared to the NJL interaction
term $\phi_1$. This can be realized by noting that $\phi_0\sim\lambda^{1/4}$
while $\phi_1\sim\lambda$. In the strong coupling limit $\lambda\gg 1$, for
the quark one-point function we have to retain only the NJL term. Therefore,
we obtain
\begin{equation}
(i\gamma^\mu\partial_\mu-m_q)q_i^1(x)+\frac{g^2}{N_c^2-1}\sum_\eta\int d^4y
  \Delta(x-y)\gamma^\mu\eta_\mu^aT_aq_i^1(x)
  \sum_j\left[{\bar q}_j^1(y)\gamma^\nu\eta_\nu^bT_bq_j^1(y)\right]=0.
\end{equation}
The equation turns out to be the Euler--Lagrange equation with respect to the
one-point function of the quark, obtained for the Nambu--Jona-Lasinio model
with a non-local Lagrangian~\cite{Bowler:1994ir,GomezDumm:2006vz}
\begin{eqnarray} \label{NJLeq1}
\lefteqn{{\cal L}''_{\rm NJL}
  \ =\ \sum_i\bar q_i^1(x)(i\gamma^\mu\partial_\mu-m_q)q_i^1(x)
  \strut}\nonumber\\&&\strut
  +\frac{g^2}{N_c^2-1}\sum_\eta
  \sum_i\left[\bar q_i^1(x)\gamma^\mu\eta_\mu^aT_aq_i^1(x)\right]
  \int d^4y\Delta(x-y)
  \sum_j\left[\bar q_j^1(y)\gamma^\nu\eta_\nu^bT_bq_j^1(y)\right].
\end{eqnarray}
Note that $\sum_\eta\eta_\mu^a\eta_\nu^b=\delta_{ab}g_{\mu\nu}$, where $\eta$
symbolizes the polarizations. In addition, one traces out the color degrees of
freedom with $\tr(T_aT_a)=N_cC_F$, $C_F=(N_c^2-1)/(2N_c)$, and $\sum_i
\bar q_i^1(x)\gamma^\mu q_i^1(x)=N_c\sum_i\bar\psi_i(x)\gamma^\mu\psi_i(x)$,
where $\psi_i(x)$ are spinors in Dirac and flavor space, only. This leads us
to the NJL lagrangian
\begin{eqnarray}
\lefteqn{{\cal L}'_{\rm NJL}\ =\ \sum_i\bar\psi_i(x)
  (i\gamma^\mu\partial_\mu-m_q)q_i(x)\strut}\nonumber\\&&\strut
  +\frac{N_cg^2}2\sum_i\left[\bar\psi_i(x)\gamma^\mu\psi_i(x)\right]
  \int d^4y\Delta(x-y)\sum_j\left[\bar\psi_j(y)\gamma_\mu\psi_j(y)\right].
\end{eqnarray}
The Fierz rearrangement of the quark fields yields
\begin{eqnarray}\label{fierz}
\lefteqn{{\cal L}'_{\rm NJL}\ =\ \sum_i\bar\psi_i(x)
  (i\gamma^\mu\partial_\mu-m_q)\psi_i(x)\strut}\nonumber\\&&\strut
  +\frac{N_cg^2}2\int d^4y\Delta(x-y)\sum_{i,j}\bar\psi_i(x)\psi_j(y)
  \bar\psi_j(y)\psi_i(x)\strut\nonumber\\&&\strut
  +\frac{N_cg^2}2\int d^4y\Delta(x-y)\sum_{i,j}\bar\psi_i(x)i\gamma_5\psi_j(y)
  \bar\psi_j(y)i\gamma_5\psi_i(x)\strut\nonumber\\&&\strut
  -\frac{N_cg^2}4\int d^4y\Delta(x-y)\sum_{i,j}\bar\psi_i(x)\gamma^\mu
  \psi_j(y)\bar\psi_j(y)\gamma_\mu \psi_i(x)\strut\nonumber\\&&\strut
  -\frac{N_cg^2}4\int d^4y\Delta(x-y)\sum_{i,j}
  \bar\psi_i(x)\gamma^\mu\gamma_5\psi_j(y)
  \bar\psi_j(y)\gamma_\mu\gamma_5\psi_i(x).
\end{eqnarray}

\subsection{Bosonization}
Let $\Gamma_\alpha$ be a set of Dirac and flavor matrices containing not only
the Dirac structures $1$, $i\gamma_5$, $\gamma_\mu$ and $\gamma_\mu\gamma_5$
from the Fierz rearrangement but also the flavor matrices $\bbbone$ and
$\frac12\lambda_\alpha$ relating quarks of equal and different flavor $i$ and
$j$ in adjoint representation. $\Gamma_\alpha$ obeys the conjugation rule
$\gamma^0\Gamma_\alpha^\dagger\gamma^0=\Gamma_\alpha$, where $\alpha$ denotes
the components of the adjoint flavor representation. Accordingly, the spinor
$\psi(x)$ spans over all these spaces. The most prominent degrees of freedom
are the scalar--isoscalar and pseudoscalar--isovector degrees which can
formally be combined as four vector. As the coefficients of these two
contributions are the same, one can reinterpret the sum over these $1+3=4$
degrees of freedom as a sum over four-vector components. The next step is to
apply the bosonization procedure exemplified in Ref.~\cite{Hell:2008cc} by
adding scalar--isoscalar and pseudoscalar--isovector mesonic fields as
auxiliary fields $M_\alpha(w)=(\sigma(w);\vec\pi(w))$ at an intermediate
space-time location $w=(x+y)/2$, coupled to the nonlocal fermionic currents.
The result of the Fierz rearrangement can be expressed as NJL action
\begin{eqnarray}
\lefteqn{{\cal S}_{\rm NJL}\ =\ -\frac{N_cg^2}{2G^2}\int d^4z\Delta(z)
  \int d^4wM_\alpha^*(w)M^\alpha(w)\strut}\\&&\strut
  +\int d^4x\Bigg[\bar\psi(x)(i\gamma^\mu\partial_\mu-m_q)\psi(x)
  +\frac{N_cg^2}2\int d^4y\Delta(x-y)\bar\psi(x)\Gamma_\alpha\psi(y)
  \bar\psi(y)\Gamma^\alpha\psi(x)\Bigg]\nonumber
\end{eqnarray}
 ($G=2\int d^4z\Delta(z)$). By performing a nonlocal functional shift
\begin{equation}
M_\alpha\pfrac{x+y}2\to M_\alpha\pfrac{x+y}2+G\bar\psi(x)\Gamma_\alpha\psi(y),
\end{equation}
the nonlocal quartic fermionic interaction can be removed. Instead, the
fermion field starts to interact nonlocally with the mesonic fields,
\begin{eqnarray}
{\cal S}_{\rm NJL}&=&-\frac{N_cg^2}{2G^2}\int d^4z\Delta(z)
  \int d^4wM_\alpha^*(w)M^\alpha(w)+\int d^4x\bar\psi(x)
  (i\gamma^\mu\partial_\mu-m_q)\psi(x)\strut\\&&\strut
  -\frac{N_cg^2}{2G}\int d^4x\int d^4y\bar\psi(x)\Delta(x-y)
  \left(M_\alpha\pfrac{x+y}2+M_\alpha^*\pfrac{x+y}2\right)\Gamma^\alpha\psi(y).
  \nonumber
\end{eqnarray}
After Fourier transform, in momentum space one obtains
\begin{eqnarray}
{\cal S}_{\rm NJL}&=&-\frac{N_cg^2}{4G}\int\frac{d^4q}{(2\pi)^4}
  \tilde M_\alpha^*(q)\tilde M^\alpha(q)+\int\frac{d^4p}{(2\pi)^4}
  \bar{\tilde\psi}(p)(\slp-m_q)\tilde\psi(p)\strut\\&&\strut
  -\frac{N_cg^2}{2G}\int\frac{d^4p}{(2\pi)^4}\int\frac{d^4p'}{(2\pi)^4}
  \bar{\tilde\psi}(p)\tilde\Delta\pfrac{p+p'}2\left(\tilde M_\alpha(p-p')
  +\tilde M_\alpha^*(p-p')\right)\Gamma^\alpha\tilde\psi(p').\nonumber
\end{eqnarray}
where the symbols with tilde are used for the Fourier transformed quantities.
The final step in the bosonization is to integrate out the fermionic fields,
in the general case leading to~\cite{Hell:2008cc}
\begin{eqnarray}
\lefteqn{{\cal S}_{\rm bos}\ =\ -\frac{N_cg^2}{4G}\int\frac{d^4q}{(2\pi)^4}
  \tilde M_\alpha^*(q)\tilde M^\alpha(q)\strut}\\&&\strut
  -\ln\det\left[(2\pi)^4\delta^4(p-p')(\slp-m_q)-\frac{N_cg^2}{2G}
  \tilde\Delta\pfrac{p+p'}2\left(\tilde M_\alpha(p-p')+\tilde M_\alpha^*(p-p')
  \right)\Gamma^\alpha\right],\nonumber
\end{eqnarray}
where $\det$ denotes the direct product of a functional and an analytical
determinant, the former in the Fock space transition between space-time
points $x$ and $y$, the latter in the Dirac and flavor indices.

\subsection{Mean field approximation}
Expanding the bosonic fields $\sigma(x)=\bar\sigma+\delta\sigma(x)$ and
$\vec\pi(x)=\delta\vec\pi(x)$ about the vacuum expectation value
$\bar\sigma=\langle\sigma\rangle$, the zeroth order expansion coefficient is
the mean field approximation, leading to the simplified NJL action
\begin{eqnarray}
{\cal S}_{\rm NJL}&=&-\frac{N_cg^2\bar\sigma^2}{4G}V^{(4)}+\int d^4x
  \bar\psi(x)(i\gamma^\mu\partial_\mu-m_q)\psi(x)\strut\nonumber\\&&\strut
  -\frac{N_cg^2\bar\sigma}G\int d^4x\int d^4y\bar\psi(x)\Delta(x-y)\psi(y).
\end{eqnarray}
After Fourier transform, in momentum space one has
\begin{eqnarray}
{\cal S}_{\rm NJL}&=&-\frac{N_cg^2\bar\sigma^2}{4G}V^{(4)}
  +\int\frac{d^4p}{(2\pi)^4}\bar{\tilde\psi}(p)(\slp-m_q)\tilde\psi(p)
  -\frac{N_cg^2\bar\sigma}{G}\int\frac{d^4p}{(2\pi)^4}\bar{\tilde\psi}(p)
  \tilde\Delta(p)\tilde\psi(p)\ =\nonumber\\
  &=&-\frac{N_cg^2\bar\sigma^2}{4G}V^{(4)}
  +\int\frac{d^4p}{(2\pi)^4}\bar{\tilde\psi}(p)(\slp-M_q(p))\tilde\psi(p)
\end{eqnarray}
with the unit space-time volume $V^{(4)}$, where ($G=2\tilde\Delta(0)$)
\begin{equation}\label{muq1}
M_q(p)=m_q+\frac{N_cg^2}G\tilde\Delta(p)\bar\sigma
  =m_q+\frac{N_cg^2\tilde\Delta(p)}{2\tilde\Delta(0)}\bar\sigma
\end{equation}
is the dynamical mass of the quark. The bosonization leads to
\begin{equation}
\frac{{\cal S}_{\rm bos}}{V^{(4)}}=-\frac{N_cg^2\bar\sigma^2}{4G}
  -\int\frac{d^4p}{(2\pi)^4}\ln\det(\slp-M_q(p)).
\end{equation}
On the other hand, one has $\ln\det(\slp-M_q(p))=\tr\ln(\slp-M_q(p))
=\frac124N_f\ln\left(p^2-M_q^2(p)\right)$. The quantity $\bar\sigma$ can be
determined by variation of the action ${\cal S}_{\rm bos}$ with respect to
this quantity. Taking into account the dependence of $M_q(p)$ on
$\bar\sigma$, one obtains
\begin{equation}
0=-\frac{N_cg^2\bar\sigma}{2G}+2N_f\int\frac{d^4p}{(2\pi)^4}
  \frac{2M_q(p)}{p^2-M_q^2(p)}\frac{N_cg^2}G\tilde\Delta(p)
  \quad\Rightarrow\quad\bar\sigma=8N_f\int\frac{d^4p}{(2\pi)^4}
  \frac{\tilde\Delta(p)M_q(p)}{p^2-M_q^2(p)}.
\end{equation}
Finally, this result can be re-inserted to Eq.~(\ref{muq1}) to obtain the
dynamical mass equation
\begin{equation}\label{muq2}
M_q(p)=m_q+4N_fN_cg^2\frac{\tilde\Delta(p)}{\tilde\Delta(0)}
\int\frac{d^4p'}{(2\pi)^4}\frac{\tilde\Delta(p')M_q(p')}{p^{\prime2}
  -M_q^2(p')}.
\end{equation}
A similar gap equation for the $g-2$ problem was shown in Ref.~\cite{Dorokhov:2016mxa}. In this article, however, we derived the gap equation directly from the QCD Lagrangian.

\section{Solving the gap equation}
At this point we can insert $\tilde\Delta(p)$ from Eq.~(\ref{Delta0mom}) into
Eq.~(\ref{muq2}) in order to obtain the gap equation for the dynamical quark
mass -- or to be more precise the couple of gap equations, if taking into
accoung Eq.~(\ref{m2}) as well. However, in order to make the calculation
feasible, we recognize that the dependence on the mass $m$ of the scalar field
is subdominant, and this mass can be neglected compared to the mass of the
quark. For $m=0$ one has $\kappa=-1$, $\varphi(\kappa=-1)=(1-i)\pi$ and
\begin{equation}
\tilde\Delta(p)=\sum_{n=0}^\infty\frac{iB_n}{p^2-m_n^2+i\epsilon},\quad
  B_n=\frac{(2n+1)^2\pi^3}{4K(-1)^3}\ \frac{e^{-(n+1/2)\pi}}{1+e^{-(2n+1)\pi}}.
\end{equation}
$m_n=(2n+1)\sqrt{2p^2}/2K(-1)=(2n+1)m_0$ is the glue ball spectrum, with the
ground state given by $m_0=m_G(-1)=\sqrt{2p^2}/2K(-1)$ and $K(z)$ is the complete elliptic integral of the first kind. As a further
simplification we calculate the dynamical quark mass at zero momentum, $p=0$.
In this case we obtain
\begin{equation}\label{Mqeq}
M_q=m_q+4N_fN_cg^2\sum_{n=0}^\infty\int\frac{d^4p}{(2\pi)^4}
  \frac{B_n}{p^2+m_n^2}\frac{M_q}{p^2+M_q^2}.
\end{equation}
where we have performed a Wick rotation to the Euclidean domain. As this
integral is UV singular, we integrate the momentum up to a cut $\Lambda$ to
obtain
\begin{eqnarray}
\lefteqn{\int^\Lambda\frac{d^4p}{(2\pi)^4}\frac{B_n}{p^2+m_n^2}
  \frac{M_q}{p^2+M_q^2}\ =\ \frac{\pi^2}{(2\pi)^4}\int_0^{\Lambda^2}
  \frac{B_nM_qp^2dp^2}{(p^2+m_n^2)(p^2+M_q^2)}\ =}\nonumber\\
  &=&\frac1{(4\pi)^2(m_n^2-M_q^2)}\left[
  m_n^2\ln\left(1+\frac{\Lambda^2}{m_n^2}\right)
  -M_q^2\ln\left(1+\frac{\Lambda^2}{M_q^2}\right)\right]\ =\nonumber\\
  &=&\frac1{(4\pi)^2((2n+1)x^2-y^2)}\left[(2n+1)^2x^2
  \ln\left(1+\frac1{(2n+1)^2x^2}\right)
  -y^2\ln\left(1+\frac1{y^2}\right)\right],\qquad
\end{eqnarray}
where we have used the dimensionless quantities $x=m_0/\Lambda$ and
$y=M_q/\Lambda$, assuming that $M_q\ll\Lambda$. Reinserting into
Eq.~(\ref{Mqeq}) leads to the gap equation
\begin{equation}\label{eq:Meff}
y=\frac{m_q}{\Lambda}+\kappa\alpha_s\sum_{n=0}^\infty
  \frac{B_ny}{(2n+1)^2x^2-y^2}\left[(2n+1)^2x^2\ln\left(1
  +\frac{1}{(2n+1)^2x^2}\right)-y^2\ln\left(1+\frac{1}{y^2}\right)\right],
\end{equation}
where $\kappa=N_fN_c/\pi$ and $\alpha_s=g^2/4\pi$. We note that the cut-off
completely disappeared except for the ratio $m_q/\Lambda$ that, for the light
quarks, is negligible small.

For the QCD cut-off $\Lambda=1\GeV$, the average mass of the $u$ and $d$
quarks is taken to be $m_q=0.003415(48)\GeV$~\cite{PDG}. The ground state of
the glue ball spectrum is given by the $f_0(500)$ resonance, measured as
$m_0=0.512(15)\GeV$~\cite{Ablikim:2016frj}. Using $N_c=3$, $N_f=6$ and
$\alpha_s(3.1\GeV)=0.256506$ we obtain $M_q=0.427(29)\GeV$. 

\section{Hadronic vacuum polarization}

Inspired by the approach in Ref.~\cite{deRafael:1993za}, next we will
evaluate the contribution to the hadronic vacuum polarization, assuming
that a NJL approximation holds~\cite{Frasca:2013kka,Frasca:2019ysi}. Looking
at the Fierz decomposition as shown in Eqn.~(\ref{fierz}), one obtains
\begin{equation}
G_S=G_P=\pi\alpha G,\qquad G_V=G_A=-\frac12\pi\alpha G,
\end{equation}
where
\begin{equation}
G=2\tilde\Delta(0)=-\sum_{n=0}^\infty\frac{B_n}{(2n+1)^2m_0^2},
\end{equation}
which agrees well with the analysis in the preceding section, provided we
evaluate the gap equation as in Eq.~(\ref{eq:Meff}).
Using Ref.~\cite{deRafael:1993za}, we evaluate
\begin{equation}
a_\mu=\pfrac\alpha\pi^2m_\mu^2\frac{4\pi^2}3P_1.
\end{equation}
The coefficient $P_1$ determines the contribution called ``had 1a''. It is
defined by
\begin{equation}
P_1=-\frac{\partial\Pi_R^{(H)}(Q^2)}{\partial Q^2}\Bigg|_{Q^2=0}
\end{equation}
where $\Pi_R^{(H)}(Q^2)=\frac23\left(\Pi_V^{(1)}(Q^2)-\Pi_V^{(1)}(0)\right)$,
\begin{equation}
\Pi_V^{(1)}(Q^2)=\frac{\bar\Pi_V^{(1)}(Q^2)}{1+Q^2(8\pi^2G_V/N_c\Lambda_\chi)
  \bar\Pi_V^{(1)}(Q^2)},
\end{equation}
and
\begin{equation}
\bar\Pi_V^{(1)}(Q^2)=\frac{N_c}{2\pi^2}\int_0^1dy\,y(1-y)
  \Gamma\left(0,\frac{M_q^2+Q^2y(1-y)}{\Lambda_\chi^2}\right),\qquad
\Gamma(n,\eps)=\int_\eps^\infty\frac{dz}ze^{-z}z^n.
\end{equation}
$\Gamma(n,\eps)$ is the incomplete gamma function, but $\Gamma(1,\eps)$ is
an analytic expression,
\begin{equation}
\Gamma(1,\eps)=\int_\eps^\infty e^{-z}dz=\Big[-e^{-z}\Big]_{z=\eps}^\infty
  =e^{-\eps}.
\end{equation}
Using these formulas, we obtain
\begin{eqnarray}
P_1&=&-\frac{\partial\Pi_R^{(H)}(Q^2)}{\partial Q^2}\Bigg|_{Q^2=0}
  \ =\ -\frac23\frac{\partial\Pi_V^{(1)}(Q^2)}{\partial Q^2}\Bigg|_{Q^2=0}
  \ =\nonumber\\
  &=&-\frac23\Bigg[\frac1{1+Q^2(8\pi^2G_V/N_c\Lambda_\chi^2)
  \bar\Pi_V^{(1)}(Q^2)}\frac{\partial\bar\Pi_V^{(1)}(Q^2)}{\partial Q^2}
  \strut\nonumber\\&&\strut\qquad
  -\frac{\bar\Pi_V^{(1)}(Q^2)}{\left(1+Q^2(8\pi^2G_V/N_c\Lambda_\chi^2)
  \bar\Pi_V^{(1)}(Q^2)\right)^2}\frac{8\pi^2G_V}{N_c\Lambda_\chi^2}
  \left(\bar\Pi_V^{(1)}(Q^2)+Q^2\frac{\partial\bar\Pi_V^{(1)}(Q^2)}{\partial
  Q^2}\right)\Bigg]_{Q^2=0}\ =\nonumber\\
  &=&-\frac23\left[\frac{\partial\bar\Pi_V^{(1)}(Q^2)}{\partial Q^2}
  -\frac{8\pi^2G_V}{N_c\Lambda_\chi^2}\bar\Pi_V^{(1)}(Q^2)^2\right]_{Q^2=0}
  \ =\nonumber\\
  &=&-\frac23\Bigg[-\frac{N_c}{2\pi^2}\int_0^1dy
  \frac{y^2(1-y)^2}{\Lambda_\chi^2}\frac{\Lambda_\chi^2}{M_q^2+Q^2y(1-y)}
  e^{-(M_q^2+Q^2y(1-y))/\Lambda_\chi^2}\Bigg|_{Q^2=0}
  \strut\nonumber\\&&\strut\qquad
  -\frac{8\pi^2G_V}{N_c\Lambda_\chi^2}\left(\frac{N_c}{2\pi^2}
  \int_0^1dy\,y(1-y)\Gamma\left(0,\frac{M_q^2}{\Lambda_\chi^2}\right)\right)^2
  \Bigg]\ =\nonumber\\
  &=&-\frac23\Bigg[-\frac{N_c}{60\pi^2M_q^2}\Gamma\left(1,
  \frac{M_q^2}{\Lambda_\chi^2}\right)-\frac{N_cG_V}{18\pi^2\Lambda_\chi^2}
  \Gamma\left(0,\frac{M_q^2}{\Lambda_\chi^2}\right)^2\Bigg]\ =\nonumber\\
  &=&\frac{N_c}{3\pi^2}\frac1{30M_q^2}
  \Bigg[\Gamma\left(1,\frac{M_q^2}{\Lambda_\chi^2}\right)
  +\frac{10G_VM_q^2}{3\Lambda_\chi^2}\Gamma\left(0,\frac{M_q^2}{\Lambda_\chi^2}
  \right)^2\Bigg].
\end{eqnarray}
With the values given above, for the $u$ and $d$ quarks we obtain
\begin{equation} \label{result}
a_\mu^{u,d}(\mbox{had 1a})=452(67)\cdot 10^{-10}.
\end{equation}
This result is
in close agreement with the evaluation given
in eq.(3.3) in Ref.~\cite{Keshavarzi:2018mgv} and eq.(18)
in Ref.~\cite{Colangelo:2020lcg} and Eq.(6) in Ref.~\cite{Davier:2019can}.

In order to have a clearer understanding of the meaning of this result, we present also the strange quark contribution. This will yield
\begin{equation} \label{result2}
a_\mu^{s}(\mbox{had 1a})=232(34)\cdot 10^{-10}.
\end{equation}
The overall is
\be
a_\mu^{HVP}=684(75)\cdot 10^{-10}.
\ee
The error bar is not yet competitive to decide if BSM physics is needed but nevertheless in closed agreement with the experimental value 
as obtained in \cite{Keshavarzi:2018mgv,Aoyama:2020ynm,Colangelo:2020lcg} 
from experiments in hadron physics.

Finally, we want to analyze the contribution to the error due to the choice of the 't~Hooft limit: $Ng^2$ constant and $N\rightarrow\infty$. There have been several studies on lattice to estimate the error of such an approximation (\cite{Bali:2013kia,Perez:2020vbn} and references therein). The main conclusion is that the next-to-leading order correction to any observable goes like
\be
A=A(\infty)+\frac{c_1}{N^2}+\ldots,
\ee
being $c_1=O(1)$, a numerical factor. This same pattern is seen in the spectrum of a Yang-Mills theory without quarks where, for the ground state, one sees \cite{Lucini:2013qja}
\be
\frac{m_{0^{++}}}{\sqrt{\sigma}}=3.28(8)+\frac{2.1(1.1)}{N^2},
\ee
where $\sigma$ is a mass scale proper to strong interactions and obtained by experiment. So, this can be estimated of the same magnitude as the error we obtained from QCD data at worst.

\medskip

\section{Conclusions and Outlook\label{conc}}

To summarise, using technique devised by Bender, Milton and Savage, in Ref~\cite{Bender:1999ek} the Dyson-Schwinger equations for quantum chromodynamics in differential form was
revisited. Following Ref.~\cite{Bender:1999ek}, in this article we discussed the hadronic contributions to the muon anomalous magnetic moment following NJL model as the low energy
effective theory description of QCD, as shown in Eq.~(\ref{NJLeq1}). We provided a full derivation of the HVP contribution to the anomalous magnetic moment $a=(g-2)/2$ of the muon from first principles, starting from the QCD partition function and the effective mass for the quarks as shown in Eq.~(\ref{muq2}). Our result as obtained in Eq.~(\ref{result}) is in close agreement with the Muon $g-2$ Theory Initiative~\cite{Aoyama:2020ynm}, 
%
as obtained from experimental data in
Ref.~\cite{Keshavarzi:2018mgv,Colangelo:2020lcg,Davier:2019can}. 
In doing so, we have shown a possible new analytical approach as an alternative to lattice calculations. 
%
Our approach provides a theoretical framework for the application of QCD to several other applications and the opportunity to investigate future studies model-building for BSM physics in the dark sector just by using analytical methods. The next step will be to include other quark flavors which is beyond the scope of the current manuscript. Moreover, following the same approach and using NJL model as the low energy EFT for QCD, we also can perform a complete proof of confinement in QCD in our future studies.

We hope to improve our computations in the near future to reduce the error bar significantly.

\medskip

\section{Acknowledgements\label{Ack}}
The research was supported in part by the European Regional Development Fund
under Grant No.~TK133.

\end{document}